\documentclass[twoside]{ae100prg}
\usepackage{graphicx}
\usepackage[breaklinks]{hyperref}
\usepackage{booktabs}

\begin{document}
\title{Cosmology in $f(R)$ exponential gravity}


\author{Luisa Jaime$^{1,2}$, Marcelo Salgado$^1$ and 
Leonardo Pati\~no$^2$}

\address{$^1$ Instituto de Ciencias Nucleares, Universidad Nacional
Aut\'onoma de M\'exico, A.P. 70-543, M\'exico D.F. 04510, M\'exico}
\address{$^2$ Facultad de Ciencias, Universidad Nacional
Aut\'onoma de M\'exico, A.P. 50-542, M\'exico D.F. 04510, M\'exico}

\email{marcelo@nucleares.unam.mx}

\begin{abstract}
Using an approach that treats the Ricci scalar itself as a degree of 
freedom, we analyze the cosmological evolution within an $f(R)$ model that 
has been proposed recently (exponential gravity) and that can be viable for explaining 
the accelerated expansion and other features of the Universe. 
This approach differs from the usual scalar--tensor method and, 
among other things, it spares us from dealing with unnecessary discussions 
about {\it frames}. It also leads to a simple system of equations which is particularly 
suited for a numerical analysis.
\end{abstract}

\section{Introduction}
Since the discovery of the accelerated expansion of the Universe from the observations
of supernovae Ia~\cite{Perlmutter1999,Riess1998,Amanullah2010} and its interpretation using the $\Lambda$CDM model of 
standard cosmology, a large amount of investigations have been devoted to explain the 
same phenomenon but using dark energy substances different from the cosmological constant $\Lambda$. 
One can rank the dark energy models from the most ``conservative'' to the most 
``radical'' ones. Among the former we can mention those which do not 
introduce new fields or modifications to general relativity but which consider that 
inhomogeneities in the Universe could be enough to account for the observations~\cite{inhomogeneous}. 
There are also the models that introduce new fields and perfect fluids with exotic 
equations of state within the framework of general relativity (GR) in order to avoid the 
``problems'' associated with $\Lambda$~\cite{Sahni2000,Carroll2001,Martin2012,Bianchi2010}. 
For instance, {\it quintessence}, {\it $k$--essence}, and Chaplygin gas are some of the most 
popular models of this kind. The most radical attempts to explain the accelerated expansion are perhaps those 
which propose to modify GR while keeping the hypothesis of homogeneity and isotropy as a first approximation.
Many alternate theories of gravity have been proposed to explain this phenomenon 
as well as those related with the dark matter (e.g. rotation curves of galaxies). 
The modified $f(R)$ metric gravity is just one of such theories and 
maybe the most analyzed one in the last ten years, where the geometry takes care of mimicking the dark energy. 
This alternative is certainly radical since GR 
has been thoroughly tested for almost one hundred years and it has not only supported all 
the tests but in addition most of its predictions have been confirmed as well. 
Thus, the challenge of modified gravity is both to be consistent with GR tests and also to explain the phenomena they 
were called for. This is a no trivial 
task and many $f(R)$ models have failed in the attempt. The story concerning dark substances does not end 
with the accelerated expansion. The measurements of the angular distribution of cosmic background 
radiation anisotropies in the sky can also be explained by the $\Lambda$CDM model, and therefore, 
the task for the alternative models, theories or dark energy substances is even more demanding.

As we mentioned, $f(R)$ theories have been studied in detail in a recent past and it is out of the scope 
of the present article to discuss all the properties, problems and features 
associated with some of the specific models proposed before 
(see Refs. ~\cite{Nojiri2011,deFelice2010,Sotiriou2010,Capozziello2011,Clifton2011} for a review).

Our aim is to report the results of a potentially viable candidate, termed {\it exponential 
gravity}, as a model for the accelerated expansion, but using an approach that has been 
proposed recently by us~\cite{Jaime2012} and which avoids the identification with the scalar--tensor 
theories. The reason to follow this ``unorthodox'' method is because in some cases the 
scalar--tensor (ST) method can lead to ill-defined potentials, and moreover because we want to 
circumvent any possible discussion concerning the use of {\it frames} 
(Einstein vs Jordan). Debates of this sort plague the subject, some of which have only led to 
create confusion instead of shedding light.

With our technique we propose to treat the Ricci scalar itself as a degree of freedom, 
instead of using $\phi=f_R$ as in the ST method (hereafter a subindex $R$ indicates $\partial/\partial_R$). 
Our approach also spare us of inverting all the quantities depending on $R$ for treating them as functions of $\phi$. 
Moreover, we have found that in several specific applications the field equations can be recasted 
in a rather friendly way that allows us to treat them numerically or even analytically~\cite{Jaime2011,Jaime2012}. 
In the next section we present our method and apply it to the 
Friedmann-Roberson-Walker (FRW) spacetime within the scope of analyzing the cosmological evolution 
using the {\it exponential gravity} model. The analysis of other viable $f(R)$ models 
using the current approach can be seen in~\cite{Jaime2012} and in references therein
using other techniques.

\section{$f(R)$ theories, the Ricci scalar approach}
\label{fR}
The action in $f(R)$ gravity is given by:
\begin{equation}  
\label{f(R)}
S[g_{ab},{\mbox{\boldmath{$\psi$}}}] =
\!\! \int \!\! \frac{f(R)}{2\kappa} \sqrt{-g} \: d^4 x 
+ S_{\rm matt}[g_{ab}, {\mbox{\boldmath{$\psi$}}}] \; ,
\end{equation}
where  $\kappa \equiv 8\pi G_0$ (we use units where $c=1$), $f(R)$ is a sufficiently 
differentiable but otherwise {\it a priori} arbitrary function of 
the Ricci scalar $R$, and ${\mbox{\boldmath{$\psi$}}}$ represents schematically the matter 
fields. The field equation obtained from Eq.~(\ref{f(R)}) is:
\begin{equation}
\label{fieldeq1}
f_R R_{ab} -\frac{1}{2}fg_{ab} - 
\left(\nabla_a \nabla_b - g_{ab}\Box\right)f_R= \kappa T_{ab}\,\,,
\end{equation}
where $f_R$ indicates $\partial_R f$, $\Box= g^{ab}\nabla_a\nabla_b$ is the covariant 
D'Alambertian and $T_{ab}$ is the energy-momentum tensor (EMT) of matter 
associated with the ${\mbox{\boldmath{$\psi$}}}$ fields. From this equation it is 
straightforward to obtain the following equation and its trace~\cite{Jaime2011,Jaime2012}
\begin{equation}
\label{fieldeq3}
G_{ab} = 
\frac{1}{f_R}\Bigl{[} f_{RR} \nabla_a \nabla_b R + f_{RRR} (\nabla_aR)(\nabla_b R) 
- \frac{g_{ab}}{6}\Big{(} Rf_R+ f + 2\kappa T \Big{)} + \kappa T_{ab} \Bigl{]} \; ,
\end{equation}

\begin{equation}
\label{traceR}
\Box R= \frac{1}{3 f_{RR}}\Big{[}
\kappa T - 3 f_{RRR} (\nabla R)^2 + 2f- Rf_R
\Big{]} \; ,
\end{equation}
where $(\nabla R)^2:= g^{ab}(\nabla_aR)(\nabla_b R)$ and $T:= T^a_{\,\,a}$. 

The idea is then to solve simultaneously Eqs.~(\ref{fieldeq3}) and (\ref{traceR}) for the metric $g_{ab}$ and $R$ as 
a system of coupled partial differential equations. 

It is important to mention that the field equations imply that the EMT of matter alone is conserved, i.e., it satisfies 
$\nabla_a T^{ab}=0$. 

\begin{figure*}
\includegraphics[width= 5.6cm]{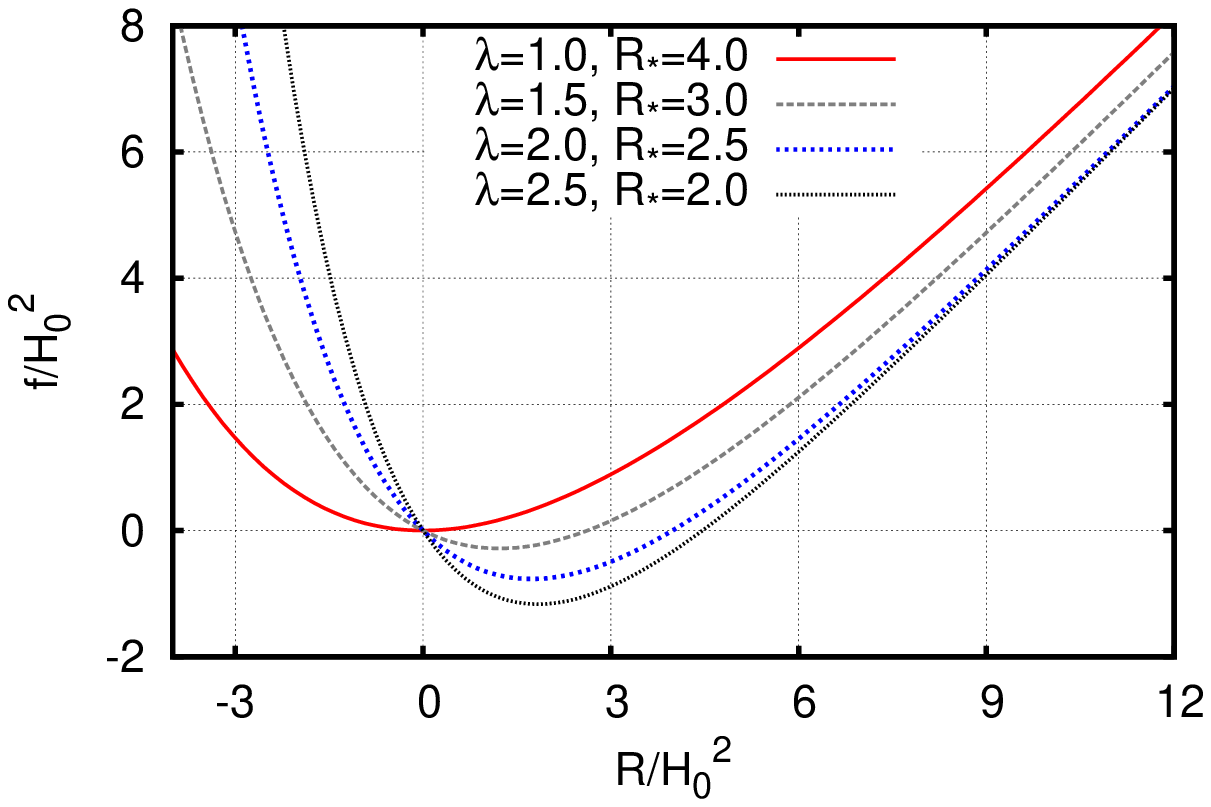}
\includegraphics[width= 5.6cm]{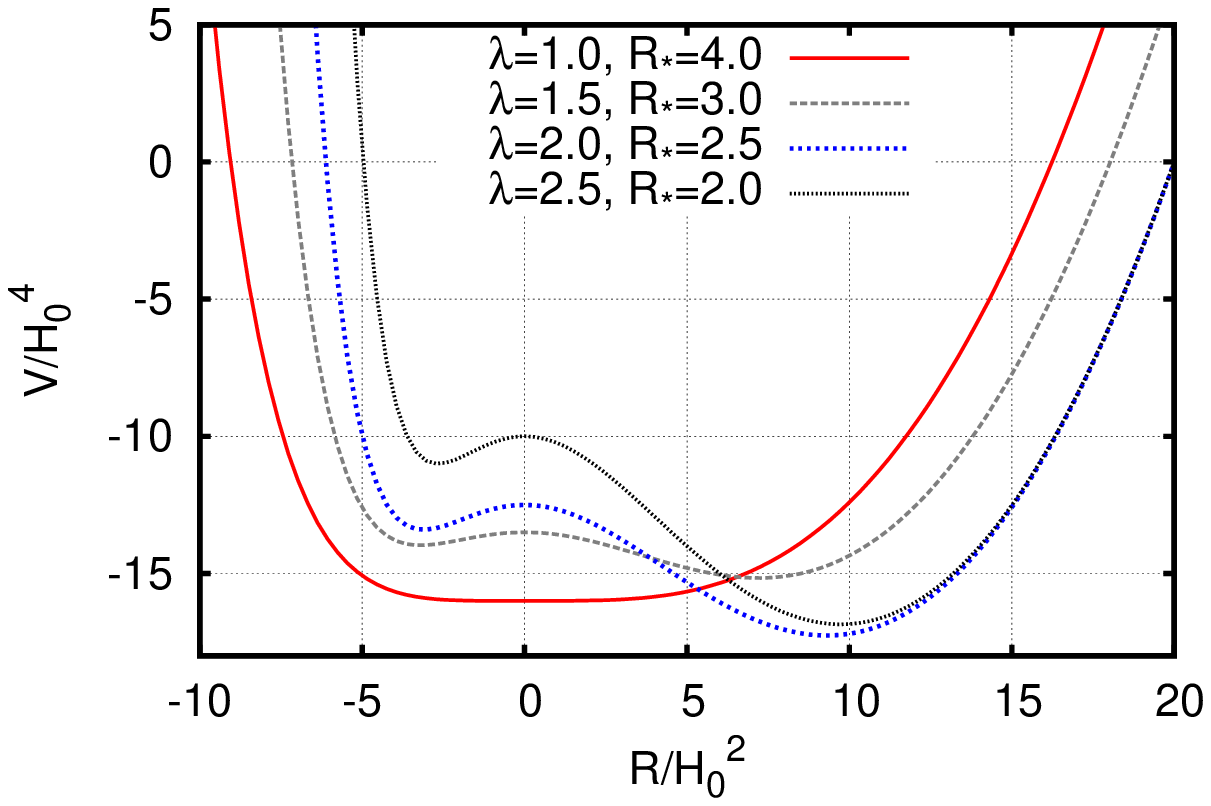}
\vskip -.2cm
\caption{(color online) $f(R)$ exponential gravity (left panel) and the potential $V(R)$ (right panel) for several values of $\lambda$.}
\label{fig:f(R)}
\end{figure*}

In this contribution we shall focus on the model $f(R)= R_*[{\tilde R} - \lambda (1- e^{-{\tilde R}})]$, referred to as 
{\it exponential gravity}, where 
${\tilde R}= R/R_*$, $\lambda$ is a positive dimensionless constant and $R_*>0$ is also a constant that fixes the built-in scale 
and which is of the order of the current Hubble parameter $H_0^2$. This kind of exponential models have been analyzed in the past by 
several authors using a different technique~\cite{Elizalde2008,Linder2009,Yang2010,Bamba2010,Elizalde2011a,Elizalde2011b}. 
Other variants of this model have also been analyzed~\cite{Zhang2006}. 
The scalar $f_R= 1- \lambda e^{-{\tilde R}}$ is positive provided $R> R_* {\rm ln} \lambda$ and this condition 
ensures $G_{\rm eff}:= G_0/f_R>0$. This latter is always satisfied in the cosmological solutions given below.
The possible de Sitter points correspond to trivial solutions $R=R_1=const.$ of Eq.~(\ref{traceR}) 
in vacuum ($T=0$) and give rise to an effective cosmological constant $\Lambda_{\rm eff}= R_1/4$ in Eq.~(\ref{fieldeq3}) 
($G_{ab}= - g_{ab} \Lambda_{\rm eff}$ in vacuum). Here $R_1>0$ is a critical point of 
the ``potential'' $V(R)$ such that $V_R(R_1)=0$ with 
$V_R:=(2f-Rf_R)/3= R_*[{\tilde R}(1+\lambda e^{-{\tilde R}}) - 2\lambda (1- e^{-{\tilde R}})]/3$, and $V_{RR}= 1- \lambda(1+{\tilde R})e^{-{\tilde R}}$. 
The ``potential'' is given by $V(R)= R_*^2[ {\tilde R}({\tilde R} - 4 \lambda) - 2\lambda({\tilde R} + 3) e^{-{\tilde R}}) ]/6$. 
For $0<\lambda \leq 1$ one can easily see that $V(R)$ has just one critical point at $R=0$ which 
is not a de Sitter point as in this case $\Lambda_{\rm eff}\equiv 0$. The point is 
a global minimum (c.f. $V_{RR}(0)=1-\lambda$). For $\lambda > 1$ there is a local maximum at $R=0$ and a global minimum at 
$R=R_1>0$ which corresponds to the actual de Sitter point that the cosmological 
solution reaches asymptotically in the future. There is also a local minimum at $R= R_2<0$, but it is an anti de Sitter point 
which is never reached as the cosmological solutions take place only in the domain $R>0$. 
The potential $V(R)$ is depicted in Figure~\ref{fig:f(R)} (right panel) where one can appreciate the critical points just described. 
In the high curvature regime ${\tilde R}\gg 1$, we have $f(R)\approx R- \lambda R_*$, and thus the model acquires 
an effective cosmological constant $\Lambda_{\rm eff}^{\infty}:= \lambda R_*/2$ (c.f. left panel of Fig.~\ref{fig:f(R)}). 
From the figure we see that for $\lambda$ sufficiently high, the de Sitter point verifies ${\tilde R}_1\gg 1$, and thus 
$R_1 \approx 2 \lambda R_*$ as it turns out from $V_R(R_1)= 0$. Therefore $\Lambda_{\rm eff}\approx \Lambda_{\rm eff}^{\infty}$. 
Finally, we stress that $f_{RR}= \lambda e^{-{\tilde R}}/R_* >0$ which ensures that no singularities are found in the equations due to 
this scalar and moreover it guarantees that given $V_{RR}(R_1)>0$ the effective mass 
$m^2:= V_{RR}/f_{RR}=(f_R-Rf_{RR})/(3f_{RR})|_{R_1}$ of the scalar mode is positive.

\section{Cosmology in $f(R)$}
We assume the spatially flat FRW metric given by:
\begin{equation}
\label{FRWmetric}
ds^2 = - dt^2  + a^2(t) \left[ dr^2 + r^2 \left(d\theta^2 + \sin^2\theta d\varphi^2\right)\right]\,\,\,.
\end{equation}
From Eqs.~(\ref{fieldeq3}) and ~(\ref{traceR}) we have,
\begin{eqnarray}
\label{traceRt}
&& \ddot R = -3H \dot R - \frac{1}{3 f_{RR}}\left[3f_{RRR} \dot R^2 + 2f- f_R R + \kappa T \right]\,\,\,,\\
\label{Hgen}
&& H^2 = \frac{\kappa}{3}\left(\rule{0mm}{0.3cm} \rho +\rho_{X}\right) \,\,\,,\\
\label{Hdotgen}
&& \dot{H}= -H^2 -\frac{\kappa}{6}\left\{\rule{0mm}{0.4cm} \rho +\rho_{X}+3\left(p_{\rm rad}+ p_{X}\right) \right\} \,\,\,. 
\end{eqnarray}
where a dot stands for $d/dt$ and $H= \dot a/a$, is the Hubble expansion. In the above equations we have included the energy 
density $\rho$ associated with matter (baryons, dark matter and radiation) as well as the GDE density $\rho_{X}$ and pressure 
$p_{X}$ given respectively by~\cite{Jaime2012}
\begin{equation}
\label{rhoX}
\rho_{X}=\frac{1}{\kappa f_{R}}\left\{\rule{0mm}{0.5cm} \frac{1}{2}\left( f_{R}R-f\right) -3f_{RR}H\dot{R} + 
\kappa \rho\left(1- f_{R}\right)\right\},
\end{equation}
\begin{equation}
\label{pressX}
p_{X}=-\frac{1}{3\kappa f_{R}}\left\lbrace \frac{1}{2}\left(f_{R}R+f \right) + 3f_{RR}H\dot{R}-\kappa\left(\rho -3 p_{\rm rad} f_R \right)
\right\rbrace \,\,\,.
\end{equation} 
Another differential equation that can be used to solve for $H$ instead of Eq.~(\ref{Hdotgen}) is given by 
$R= 6(\dot H + 2 H^2)$. This latter is no other than the Ricci scalar computed directly from the metric~(\ref{FRWmetric}). 
Equation (\ref{Hgen}) amounts to the modified Hamiltonian constraint which we use to set the 
initial data and also to monitor the accuracy of the numerical solutions at every integration step. At this regard, we stress that 
we shall not use the cosmic time $t$ but instead $\alpha= {\rm ln}(a/a_0)$ as ``time'' parameter (see Ref.~\cite{Jaime2012}), 
where $a_0$ is the present value of $a$. Notice that at the de Sitter point $R\rightarrow R_1=const.$ where $2f(R_1)= R_1 f_R(R_1)$ and 
with $\rho\rightarrow 0$, $p_{\rm rad}\rightarrow 0$ Eqs.~(\ref{rhoX}) and (\ref{pressX}), lead to $\rho_X\rightarrow \Lambda_{\rm eff}/\kappa$ and 
$p_X\rightarrow -\Lambda_{\rm eff}/\kappa$, respectively, and from Eqs.~(\ref{Hgen}) and ~(\ref{Hdotgen}), 
$H^2\rightarrow H^2_{\rm vac}= \Lambda_{\rm eff}/3= R_1/12$, and $q=-\ddot a/(aH^2)\rightarrow q_{\rm vac}= -1$.
So the main idea behind all $f(R)$ models is that as the Universe evolves, $R\rightarrow R_1$, and thus the GDE dominates and 
mimics an effective cosmological constant that allows to explain the accelerated expansion required to account for the observations.
 
The matter variables obey the conservation equation $\dot \rho_i= -3 H \left(\rho_i + p_i\right)$ for each fluid component 
(with $p_{\rm bar,DM}=0$ and $p_{\rm rad}= \rho_{\rm rad}/3$) which integrates straightforwardly and gives rise to the usual expression for 
the energy density of matter plus radiation:
$\rho= (\rho_{\rm bar}^0 + \rho_{\rm DM}^0)(a/a_0)^{-3} + \rho_{\rm rad}^0 (a/a_0)^{-4}$,
where the knotted quantities indicate their values today. The $X$--fluid variables ~(\ref{rhoX}) and (\ref{pressX}) also 
satisfy a conservation equation similar to the one above, but with an EOS $\omega_X:= p_X/\rho_X$ that evolves in cosmic time. 
Other possible inequivalent definitions of $\rho_X$, $p_X$ and $\omega_X$ have been adopted in the past, but they suffer of several 
drawbacks (see~\cite{Jaime2012} for a detailed discussion).

\begin{figure*}[t!]
\includegraphics[width= 5.6cm]{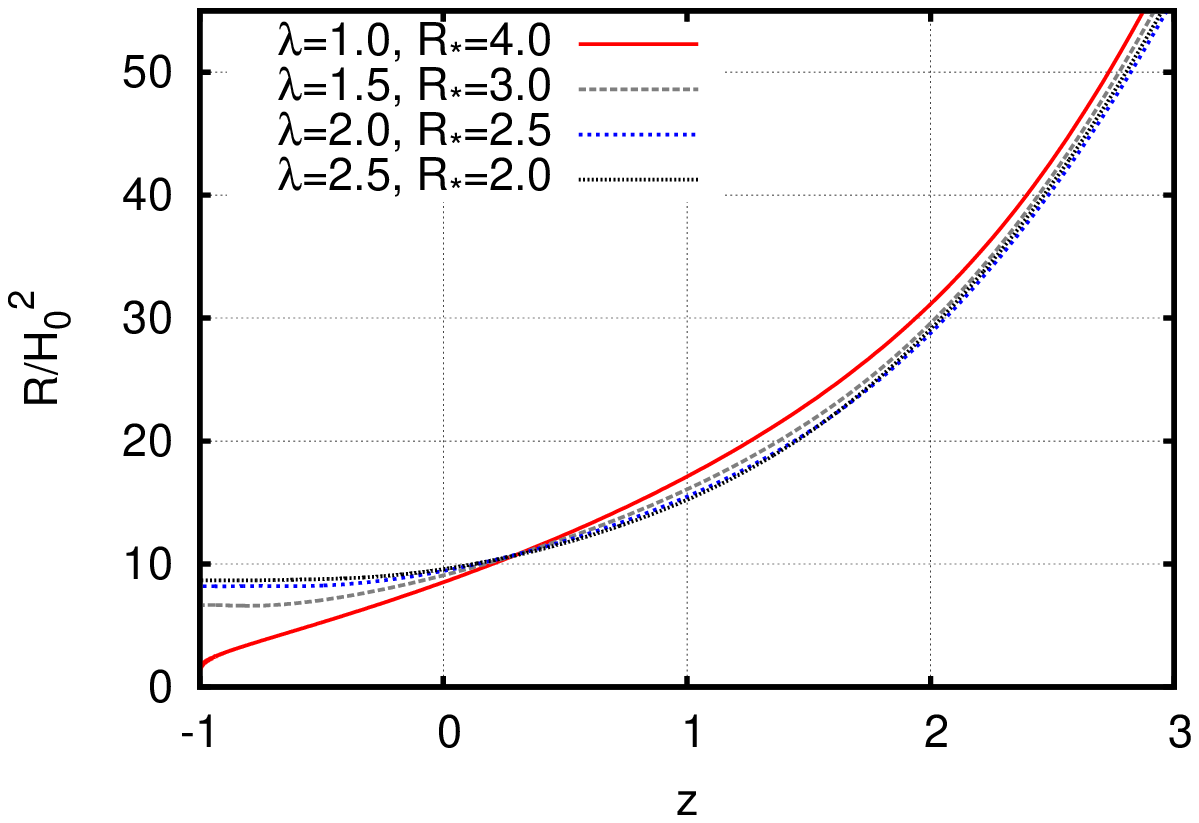}
\includegraphics[width= 5.6cm]{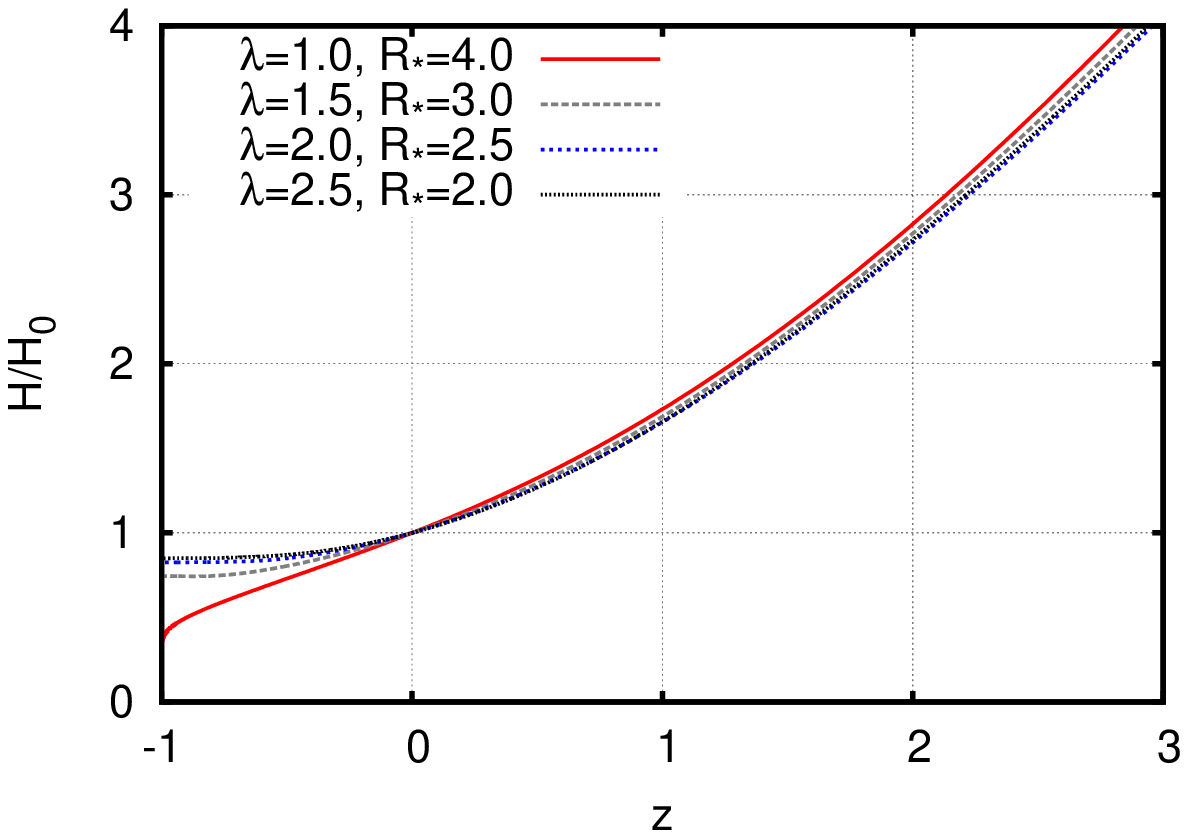}
\vskip -.2cm
\caption{(color online) Ricci scalar (left panel) and the Hubble expansion (right panel) 
for several values of $\lambda$ and $R_*$ (given in units of $H_0^2$). }
\label{fig:H-Rn}
\end{figure*}

\begin{figure*}[t]
\includegraphics[width=5.6cm]{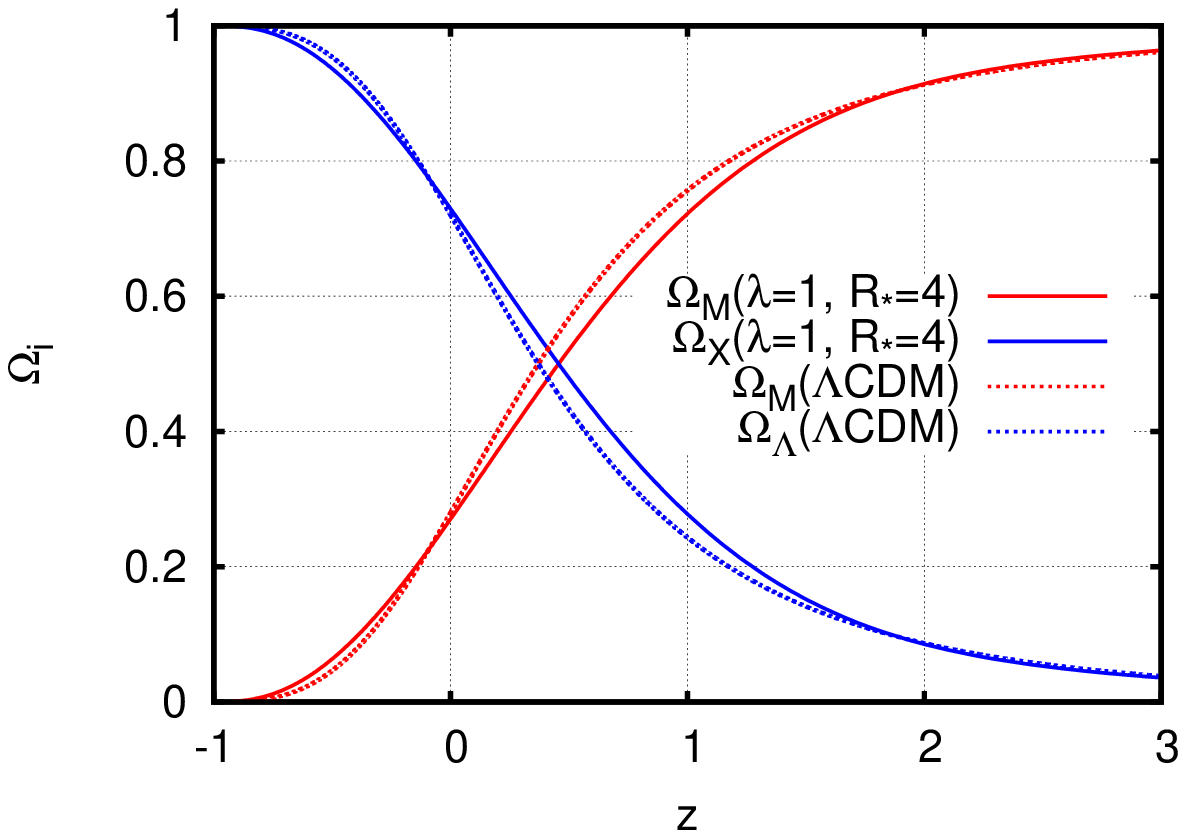}
\includegraphics[width=5.6cm]{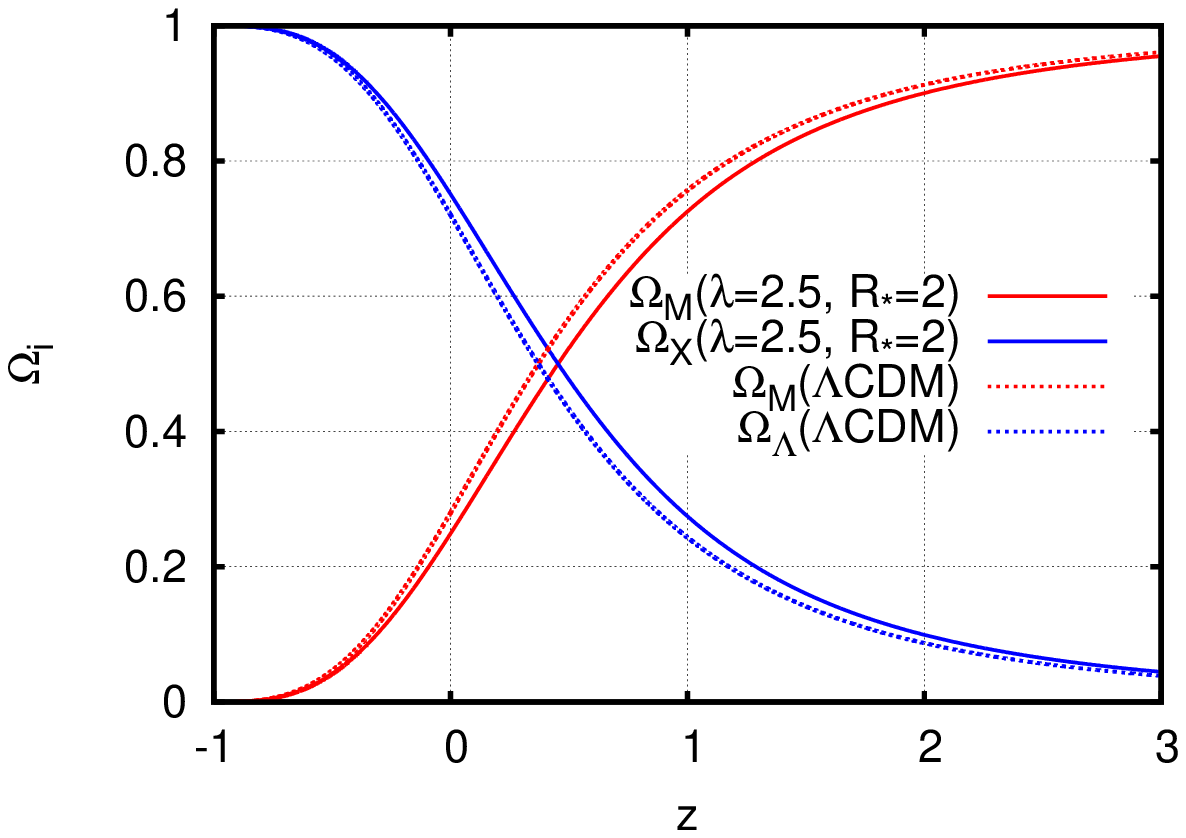}
\vskip -.4cm
 \caption{(color online) Evolution of $\Omega_{\rm matt}$ (red solid line) and 
$\Omega_X$ (blue solid line) for $\lambda= 1$ and $R_*= 4H_0^2$ (left panel) and $\lambda= 2.5$ and $R_*= 2H_0^2$ 
(right panel). For reference the corresponding quantities of the $\Lambda$CDM model are included 
in each panel (dashed lines).}
\label{fig:Omegas}
\end{figure*}

The total EOS is defined by $\omega_{\rm tot} = (p_{\rm rad}+p_{X})/(\rho +\rho_{X})$ which using Eqs.~(\ref{rhoX}) and (\ref{pressX}) yields
\begin{equation}
\label{EOSTOT}
\omega_{\rm tot} = -\frac{1}{3}\left[ \frac{\frac{1}{2}\left(f_{R}R+f \right) + 3f_{RR}H\dot{R}-\kappa\rho}
          {\frac{1}{2}\left( f_{R}R-f\right) -3f_{RR}H\dot{R} + \kappa \rho}\right] \,\,\,.
\end{equation}
This EOS allows us to track the epochs where the Universe is expanding in a decelerating or accelerating fashion. If 
$\omega_{\rm tot} < -1/3$ then $\ddot a > 0$, while $\ddot a < 0$ if $\omega_{\rm tot} > -1/3$.

\begin{figure*}[t!]
\includegraphics[width= 5.6cm]{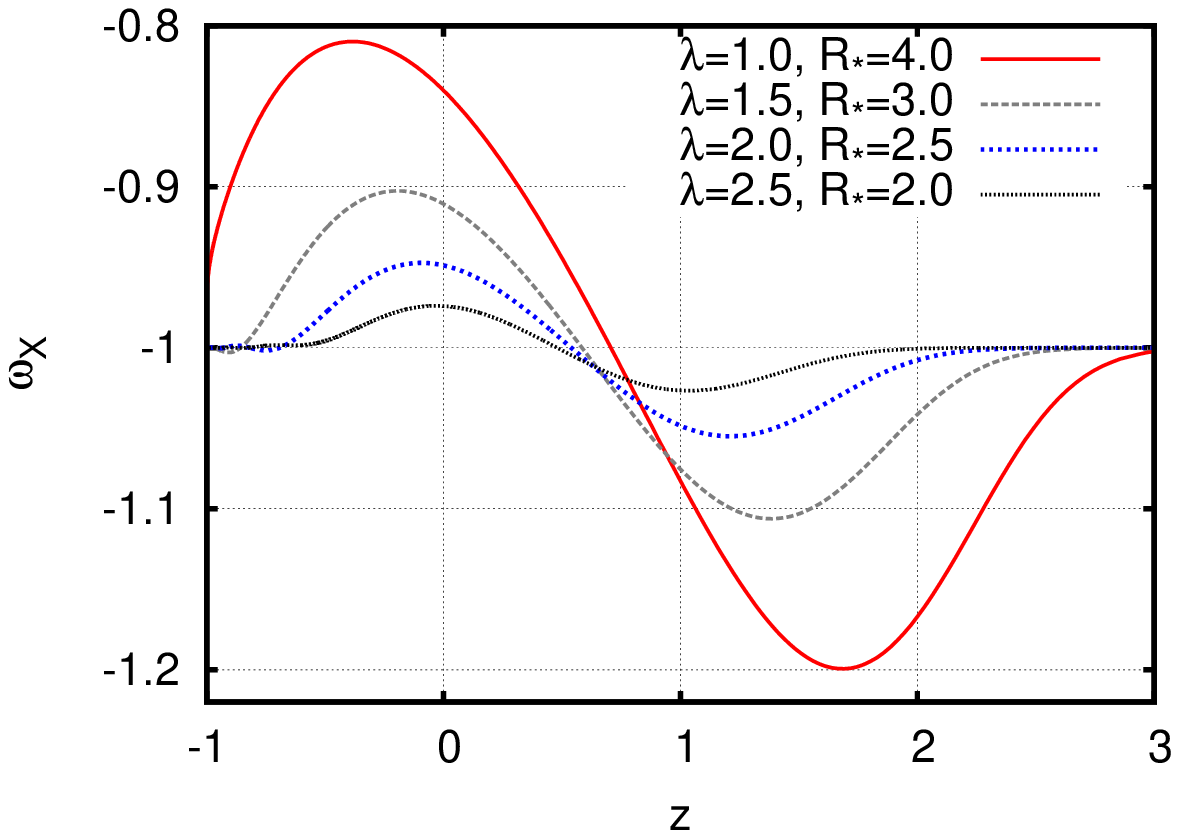} 
\includegraphics[width= 5.6cm]{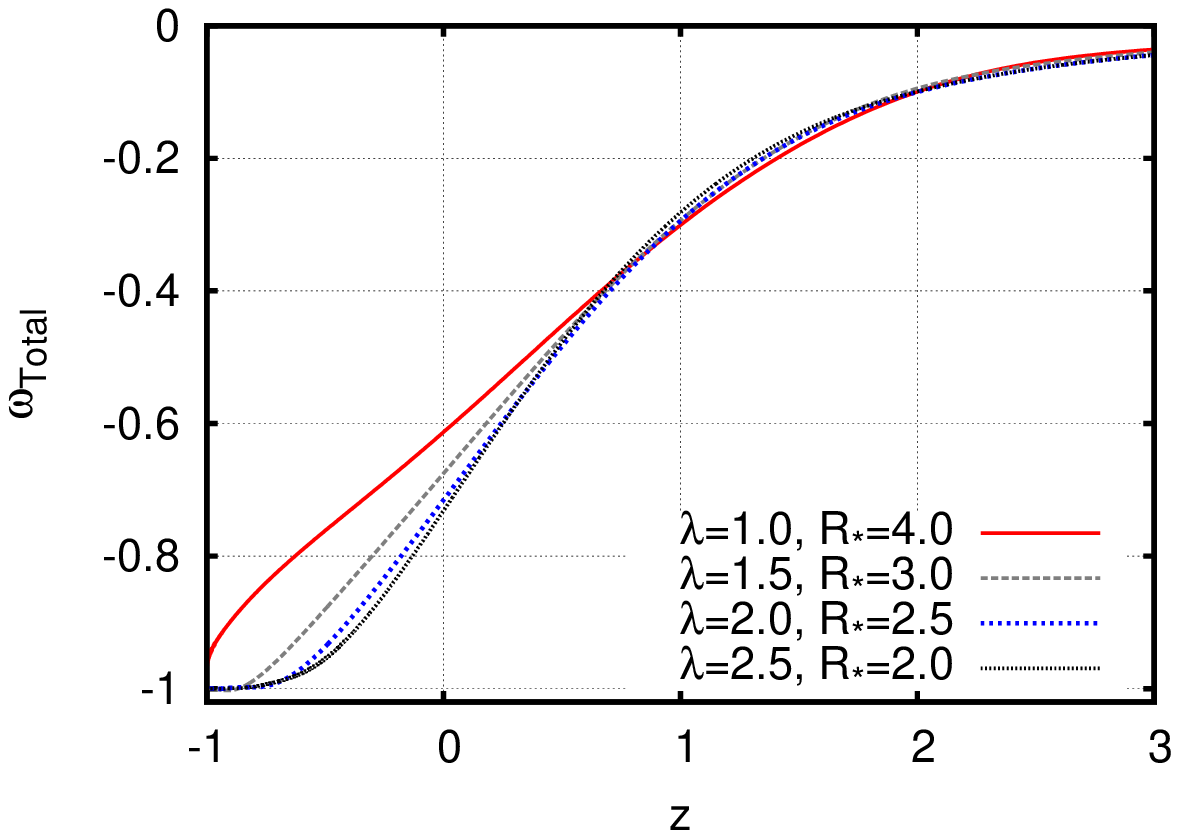}
\vskip -.2cm
\caption{(color online) The EOS $\omega_X$ (left panel) and the total EOS $\omega_{\rm tot}$ (right panel).}
\label{fig:EOS}
\end{figure*}

\section{Numerical Results and Discussion}
\begin{figure*}[t]
\includegraphics[width=5.6cm]{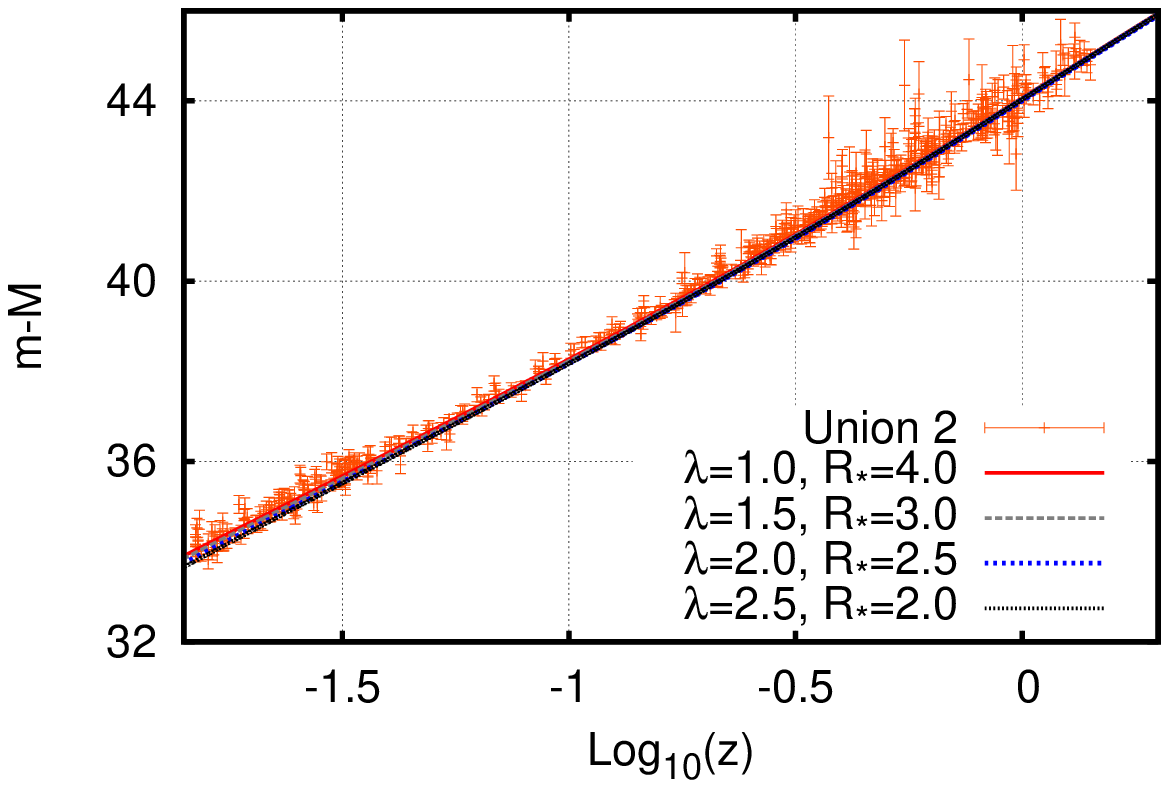}
\includegraphics[width=5.6cm]{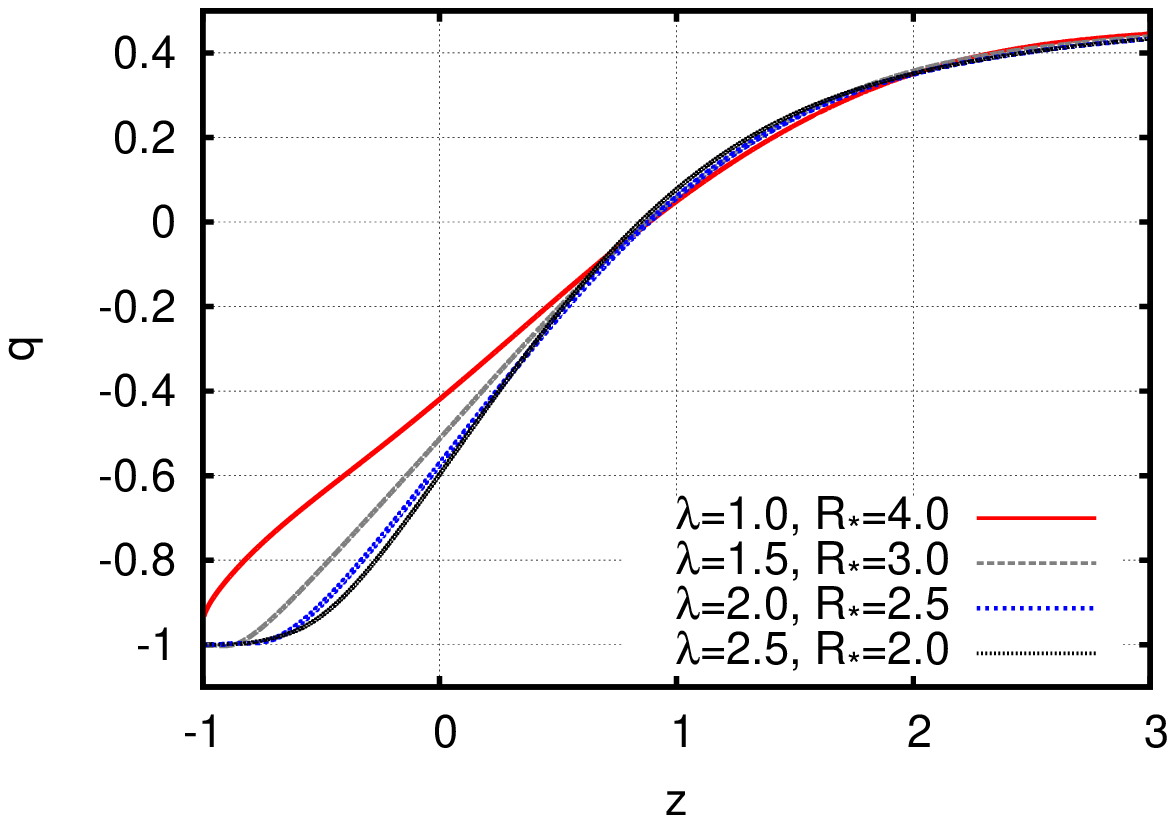}
\vskip -.3cm
 \caption{(color online) Luminous distance (left panel) compared with the Union 2 data~\cite{Amanullah2010}. Deceleration parameter 
(right panel).}
\label{fig:Lum-q}
\end{figure*}

We integrate the differential equations forward, from past to future, starting from a given $z= a_0/a -1$, where matter 
dominates, to $z\rightarrow -1$ where the GDE prevails. The initial conditions are fixed 
as in~\cite{Jaime2012}. Figure~\ref{fig:H-Rn} shows the Hubble expansion and the Ricci scalar for several values of 
$\lambda$. In all the cases where $\lambda >1$, $R$ reaches the de Sitter point at the global minimum of $V(R)$. 
For $\lambda=1$ because the potential is very flat around the global minimum at $R=0$, and also due to the friction term, 
$R$ varies very slowly as it approaches the minimum. This explains why in this case the model also mimics 
a cosmological constant. Figure~\ref{fig:Omegas} depicts the fraction of dimensionless densities 
$\Omega_i= \kappa \rho_i/(3H^2)$ which satisfy the constraint $\Omega_{\rm rad} + \Omega_{\rm matt} + \Omega_X =1$ where 
$\Omega_{\rm matt}:=  \Omega_{\rm bar}+  \Omega_{\rm DM}$. The radiation contribution, although taken into account, is very small
and cannot be appreciated from the plots. The current abundances at $z=0$ (today) match reasonably well the predicted values of the $\Lambda$CDM model 
and the exponential models show an adequate matter domination era. The EOS of GDE is plotted in Figure~\ref{fig:EOS} 
(left panel), and like in other $f(R)$ models~\cite{Jaime2012}, it oscillates around the phantom divide value $\omega_\Lambda= -1$ before reaching 
its asymptotic value as $z\rightarrow -1$. 
The total EOS depicted in Figure~\ref{fig:EOS} (right panel) shows that at higher $z$ the Universe is dominated by matter 
with $\omega_{\rm tot} \sim 0$, and then interpolates to the value $\omega_{\rm tot}= -1$ in the far future. At 
$z=0$, $\omega_{\rm tot}$ is similar to the value $\omega_{\rm tot}\sim 0.75$ predicted by the $\Lambda$CDM model. Figure~\ref{fig:Lum-q} (left panel) 
shows the (modulus) luminous distance computed as in~\cite{Jaime2012} and the deceleration parameter (right panel).

In these exponential models it is technically difficult to integrate far in the past since $f_{RR}\rightarrow 0$ exponentially. 
Since this quantity appears in the denominator of Eq.~(\ref{traceRt}), it produces large variations that affects 
the precision during the numerical integration. This is something that we had encountered in other $f(R)$ models~\cite{Jaime2012}.

The exponential model seems to be consistent with the cosmological observations and also with the Solar System~\cite{Linder2009}. 
Nevertheless, like other $f(R)$ models that look viable as well, a closer examination is required in all possible scenarios
before considering $f(R)$ theories as a serious threat to general relativity.

\section*{References}
\bibliography{salgado}

\end{document}